\pdfoutput=1
\documentclass{PoS}

\usepackage{amsmath}
\usepackage{slashed}

\title{
  $\mathcal{O}(a^2)$-improved actions for heavy quarks 
  and scaling studies on quenched lattices 
}

\ShortTitle{$\mathcal{O}(a^2)$-improved actions for heavy quarks}

\author{\speaker{Yong-Gwi Cho} \\
  Graduate School of Pure and Applied Sciences, University of Tsukuba,\\
  Tsukuba, Ibaraki 305-8571.\\
  E-mail: \email{cho@ccs.tsukuba.ac.jp}}

\author{{Shoji Hashimoto, Jun-Ichi Noaki}\\
  High Energy Accelerator Research Organization (KEK) and\\
  School of High Energy Accelerator Science, The Graduate University
  for Advanced Studies,\\
  Tsukuba 305-0801.\\
  E-mail: \email{shoji.hashimoto@kek.jp, noaki@post.kek.jp}
}

\author{Andreas J\"uttner, Marina Marinkovic\\
School of physics and astronomy, University of Southampton,
SO17 1BJ Southampton, United Kindgdom\\
E-mail: \email{a.juttner@soton.ac.uk, m.marinkovic@soton.ac.uk}}  

\abstract{
  We investigate a new class of improved relativistic fermion action on
  the lattice with a criterion to give excellent energy-momentum
  dispersion relation as well as to be consistent with tree-level
  $\mathcal{O}\left(a^{2}\right)$-improvement. Main application in mind
  is that for heavy quark for which $ma\simeq O(0.5)$.  
  We present tree-level results and a scaling study on quenched
  lattices. 
}

\FullConference{31st International Symposium on Lattice Field Theory - LATTICE 2013\\
		July 29 - August 3, 2013\\
		Mainz, Germany}

\begin{document}

\section{Introduction}
Precise non-perturbative calculation in heavy quark physics is one of
the long-standing goals of lattice QCD.
For quantities involving heavy quark, the discretization effect may
become more significant than that in the light quark sector.
With a naive order counting it appears as a power of $am$, the heavy
quark mass in unit of the lattice cutoff $1/a$, which is not much
smaller than one. 
While effective theories for heavy quarks in non-relativistic
kinematics have been developed and used on the lattice, 
a brute-force approach of taking $a$ as small as possible
would also be powerful since there is no need of additional matching of 
parameters. 
This may be combined with the Symanzik improvement of the lattice
fermion action to eliminate leading discretization effects.

The JLQCD collaboration is currently generating 2+1-flavor gauge
configurations at fine lattice spacings of $a^{-1}=$ 2.4--4.8~GeV using 
a chirally symmetric fermion formulation for light quarks 
\cite{Kaneko:2013jla}.
For the valence heavy quarks, we plan to use other fermion formulations 
that may have better scaling towards the continuum limit.

In this work we investigate some choices of the lattice fermion action
to be used for valence quarks, focusing on their discretization effect
and continuum scaling for heavy quarks.
At this initial study, we mainly consider the charm quark mass region,
and leave an extension towards heavier masses for future study.
The quantities to be studied are the energy-momentum dispersion
relation, hyperfine splitting and decay constants of the heavy-heavy mesons.
For this purpose, we are generating a series of quenched gauge
configurations that have a roughly matched physical volume 
(at about 1.6~fm) and cover a range of lattice spacings between $1/a$
= 2 and 4~GeV. 
Since these lattices do not contain sea quarks and have small physical
volume, we do not expect precise agreement with the corresponding
experimental data for the charm quark, 
but rather we are interested in their scaling towards the 
continuum limit.

The gauge configurations are generated with the tree-level
$O(a^2)$-improved Symanzik action, so far 
at $\beta$ = 4.41 and 4.66 
on $16^3\times 32$ and $24^3\times 48$ lattices, respectively.
Using the energy density expectation value after the Wilson flow, we
determine the lattice spacing with an input 
$w_0$ = 0.176(2)~fm \cite{Borsanyi:2012zs} as
$1/a$ = 1.97(2) and 2.81(3)~GeV for the two lattices.
(Note that this input value is given through the $\Omega$
baryon mass in 2+1-flavor QCD in \cite{Borsanyi:2012zs}.)
For each $\beta$ value we have analysed 100 independent gauge configurations.
In the following we mainly discuss the newly developed 
$\mathcal{O}(a^2)$-improved Brillouin fermion action, and
present our preliminary studies of the dispersion relation and
hyperfine splitting.
We also analyse the heavy-heavy decay constant calculated with the
domain-wall fermion action in the valence sector.


\section{$\mathcal{O}\left(a^{2}\right)$-improved Brillouin fermions}
\begin{figure}[tb]
  \begin{center}
    \includegraphics[width=7cm,clip=on]{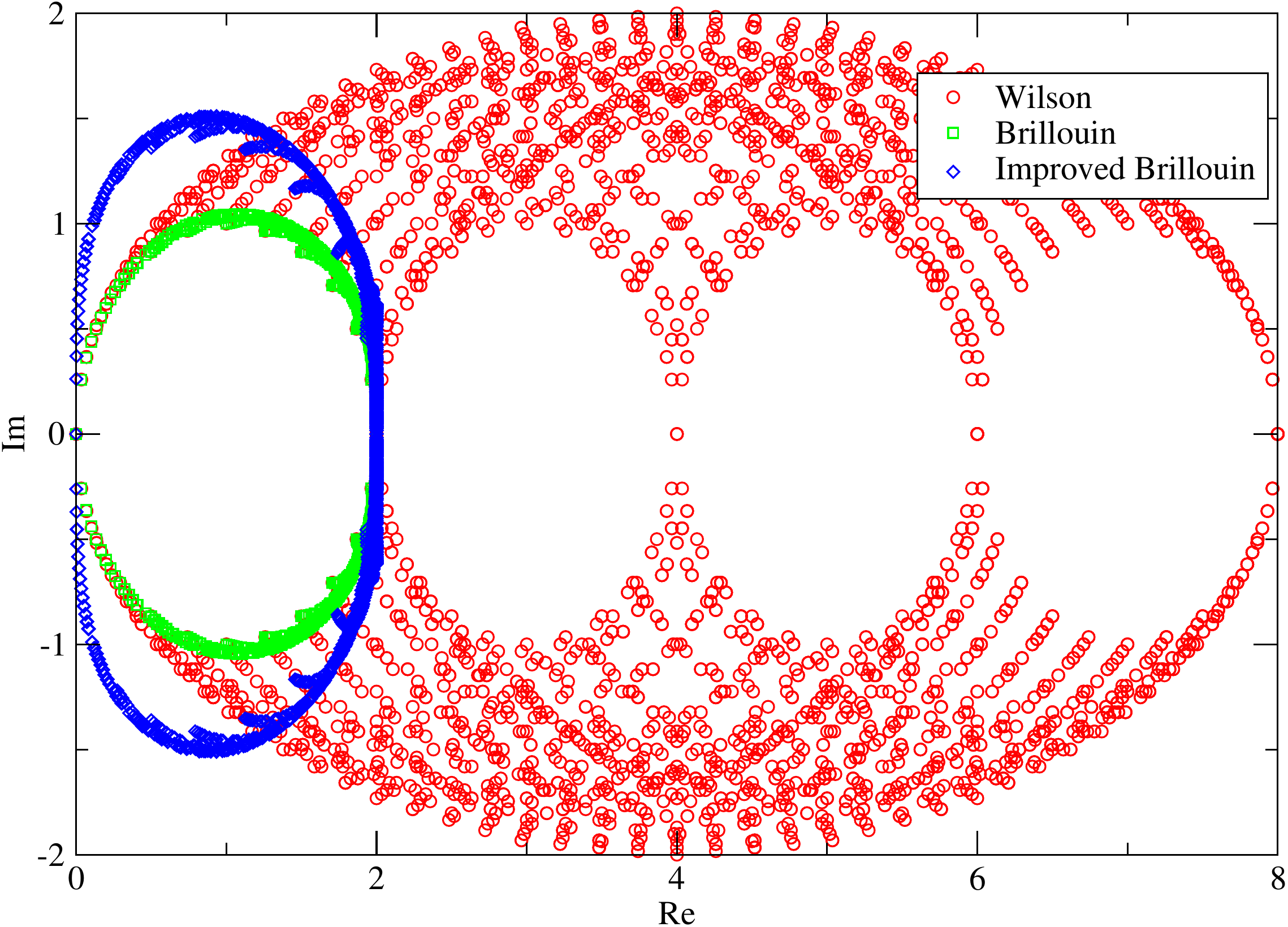}
  \end{center}
  \caption{Eigenvalue distributions of the Wilson (open circles), the
    Brillouin (open squares) and the improved Brillouin (open
    diamonds). } 
\label{fig:tree_eigen}
\end{figure}

We develop a new class of lattice fermion action which is free from
$\mathcal{O}\left(a\right)$ and $\mathcal{O}\left(a^{2}\right)$
discretization effects at the tree-level.
The action is based on the Isotropic-derivative and the Brillouin
Laplacian studied in \cite{Creutz:2010bm,Durr:2010ch}. 
The Dirac-operator is defined as 
\begin{equation}
  D^{bri}(x,y)
  =
  \sum_{\mu}\gamma_{\mu}\nabla^{iso}_{\mu}(x,y)
  -
  \frac{a}{2}
  \Delta^{bri}(x,y)+m_{0}\delta_{x,y},
\end{equation}
where $\nabla^{iso}_{\mu}(x,y)$ and $\Delta^{bri}(x,y)$ include 1-,
2-, 3- and 4-hop terms in a $3^4$ hypercube defined by
$|x_\mu-y_\mu|\le 1$, and the resulting Dirac operator is ultralocal. 
The leading discretization effect contained in 
$\nabla^{iso}_{\mu}(x,y)$
is $\mathcal{O}(a^{2})$ and is isotropic.
The Brillouin Laplacian $\Delta^{bri}(x,y)$ is designed such that all
the fifteen doublers have the same mass $2/a$ at the tree-level.
This can be seen from the eigenvalue spectrum on the complex plane as
shown in Figure~\ref{fig:tree_eigen} for the free case.
For the Wilson fermion, the spectrum has five branches, 
{\it i.e.} one corresponding to the physical (real part = 0) and the
others to doublers ($2/a$, $4/a$, $6/a$ and $8/a$).
The Brillouin operator approximately gives an unit circle centered at
(1,0), which resembles the Ginsparg-Wilson-type fermions.
This suggests that the Brillouin operator approximately satisfies the
Ginsparg-Wilson relation.

\begin{figure}[tb]
  \begin{center}
    \includegraphics[width=7cm,clip=on]{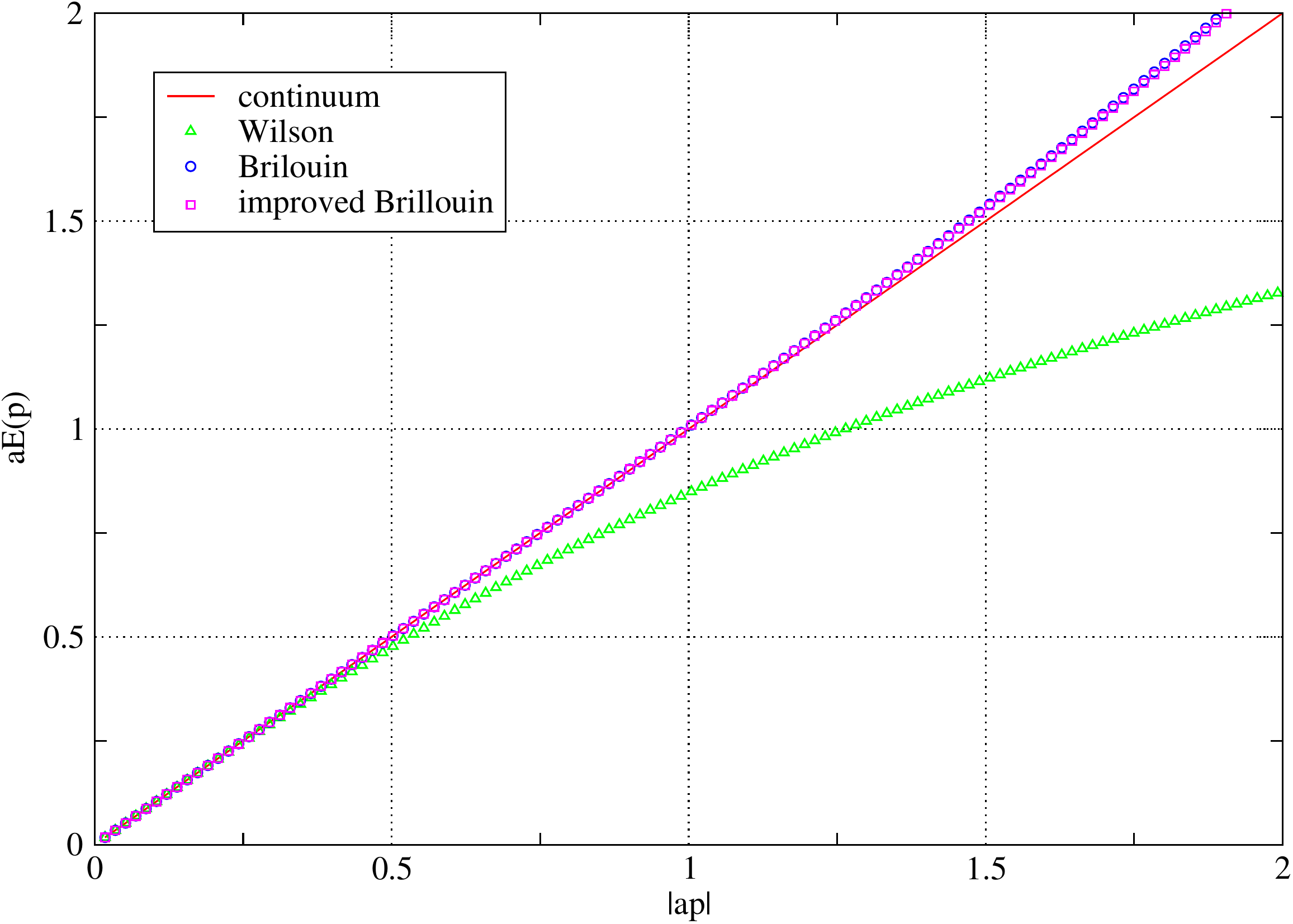}	
    \includegraphics[width=7cm,clip=on]{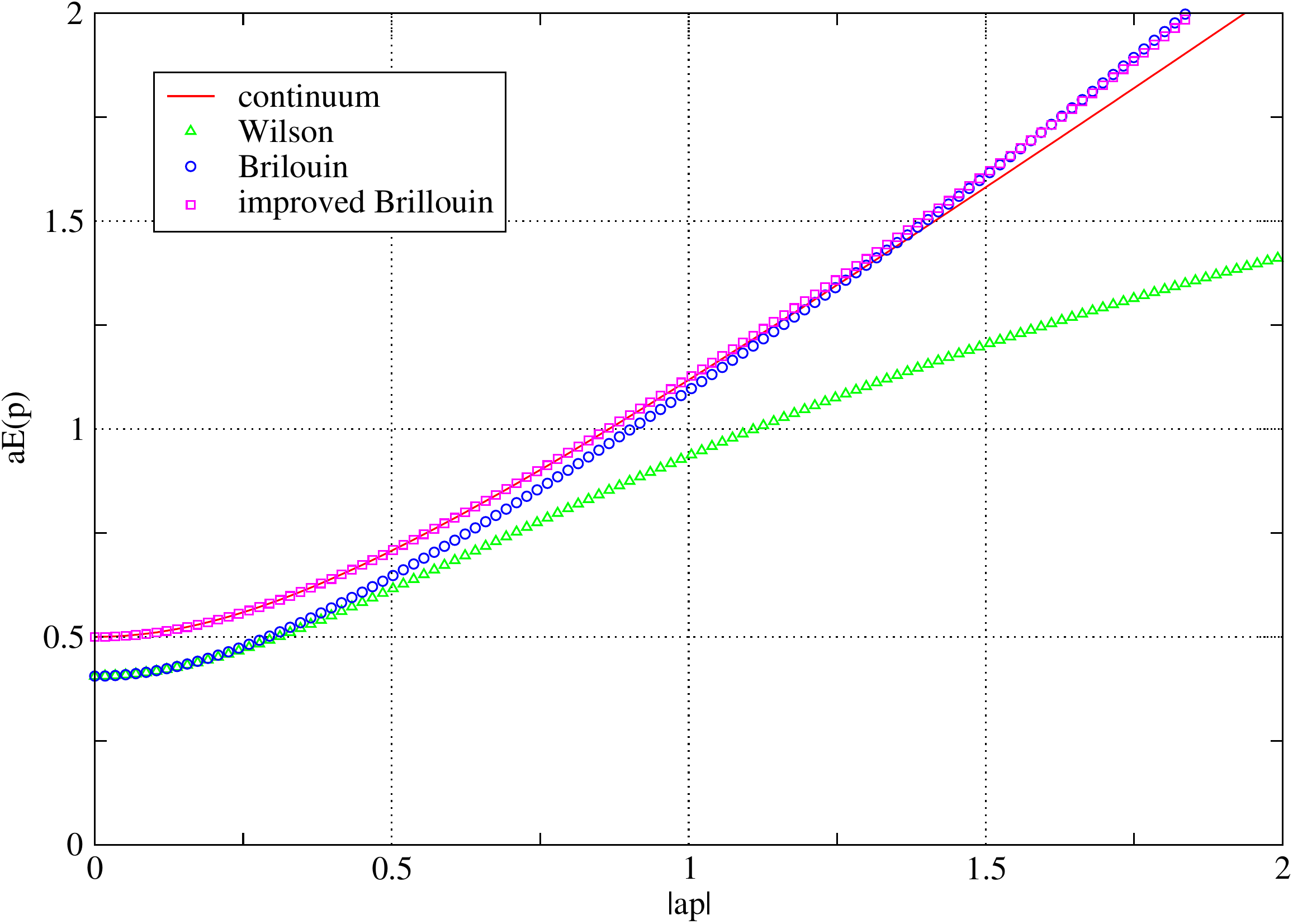}		
  \end{center}
  \caption{
    Dispersion relation calculated at the tree-level for different
    fermion formulations, {\it i.e.}
    Wilson (green), Brillouin (blue), improved Brillouin (magenta).
    The results at $ma=0.0$ (left) and $ma=0.5$ (right)
    are plotted.
  }	 
  \label{fig:dis}	
\end{figure}

With the Brillouin operator, it is found that the energy-momentum
dispersion relation calculated at the tree-level follows very
precisely that of continuum theory \cite{Durr:2010ch}, 
which is demonstrated in Figure~\ref{fig:dis}
(left: massless; right massive $ma=0.5$).
With the massless Wilson fermion, the deviation from the continuum is already seen
at $ap\sim 0.5$ in the massless case, while with the Brillouin fermion it does not start until around $ap\sim 1.5$.
This is confirmed also nonperturbatively using the dispersion relation
of mesons and baryons calculated on quenched lattices \cite{Durr:2012dw}.

In this work, we further improve the Brillouin fermion by
eliminating its leading discretization effects.
For instance, with the Brillouin operator, the relation between the
static energy $E$, defined through a pole of the free propagator, and
the bare mass $m$ is given as 
\begin{equation}
  \left(Ea\right)^{2}
  =
  (ma)^{2} - (ma)^{3}+\frac{11}{12}(ma)^{4} - \frac{5}{6}(ma)^{5} + 
  \mathcal{O}\left((ma)^{6}\right)
\end{equation}
at finite lattice spacing $a$,
and the error starts from the term of $(ma)^3$, which represents a
relative $O(a)$ effect.
Such deviation from the continuum is seen in the plot of
Figure~\ref{fig:dis} (right) at $ap=0$, where the case of $am=0.5$ is
plotted. 
Here, the Brillouin operator has a similar discretization
effect as the Wilson fermion, which gives $Ea=\ln(1+ma)$.

In order to make the Brillouin fermion consistent with the Symanzik
improvement, we eliminate the leading discretization errors
by modifying the action as
\begin{equation}
  D^{imp} =
  \sum_{\mu}\gamma_{\mu}
  \left(1-\frac{a^{2}}{12}\Delta^{bri}\right)
  \nabla^{iso}_{\mu}
  \left(1-\frac{a^{2}}{12}\Delta^{bri}\right)
  +c_{imp}a^{3}(\Delta^{bri})^{2}+m_{0}.
\end{equation}
The terms $(1-a^2\Delta^{bri}/12)$ sandwiching $\nabla_\mu^{iso}$ are
introduced to eliminate the $a^2$ errors while keeping the $\gamma_5$
hermiticity property.
The Wilson-like term is simply squared so that its effect starts from $O(a^3)$.
The relation between the energy and the bare mass becomes 
\begin{equation}
  \left(Ea\right)^{2}=(ma)^{2}+c_{imp}(ma)^{5}+\mathcal{O}\left((ma)^{6}\right),
\end{equation}%
and the leading error starts from $O(a^3)$ as expected.

For this improved Brillouin fermion action, we observe good dispersion
relation also for the massive case (see a plot on the right panel of
Figure~\ref{fig:dis}). 
The difference from the continuum is invisible below $ap\sim 1.5$.
The eigenvalues of the improved operator $D^{imp}$ no longer lie on
the unit circle as shown in Figure~\ref{fig:tree_eigen} (blue dots),
because it approaches the continuum limit which is in this case the
imaginary axis.
The blue points indeed touch the imaginary axis more closely.

Numerical implementation of the Brillouin operator is complicated
once the gauge-link is introduced, because one has to preserve
symmetries under cubic rotations for 2-, 3- and 4-hop terms.
We explicitly average over all possible paths of minimal lengths.
Computational code is implemented on the IroIro++ package
\cite{Cossu:2013ola}.

\section{Nonperturbative studies on quenched lattices}

Our scaling studies on the quenched configurations are ongoing. 
In the following we show the results for the dispersion relation and
hyperfine splitting of heavy-heavy mesons obtained with the improved
Brillouin fermion, as well as a study of heavy-heavy decay constant
using the domain-wall fermion.

For a heavy-heavy meson, we calculate an effective speed-of-light
extracted from the energy at finite momentum $\vec{p}$ as
\begin{equation}
  c_{\rm eff}^{2}(\vec{p}) = \frac{E^{2}(\vec{p})-E^{2}(\vec{0})}{\vec{p}^{2}}.
\end{equation}
The heavy quark mass is tuned until the spin-averaged 1S mass becomes 
3~GeV, and $c_{\rm eff}^2$ is calculated for the pseudo-scalar
channel.
The Wilson and improved Brillouin fermions are used on the quenched
configurations at $1/a$ = 1.97 and 2.81~GeV.
Three steps of stout smearing \cite{Morningstar:2003gk} are applied for
the gauge links. 

\begin{figure}
  \begin{center}
    \includegraphics[scale=0.3,clip=on]{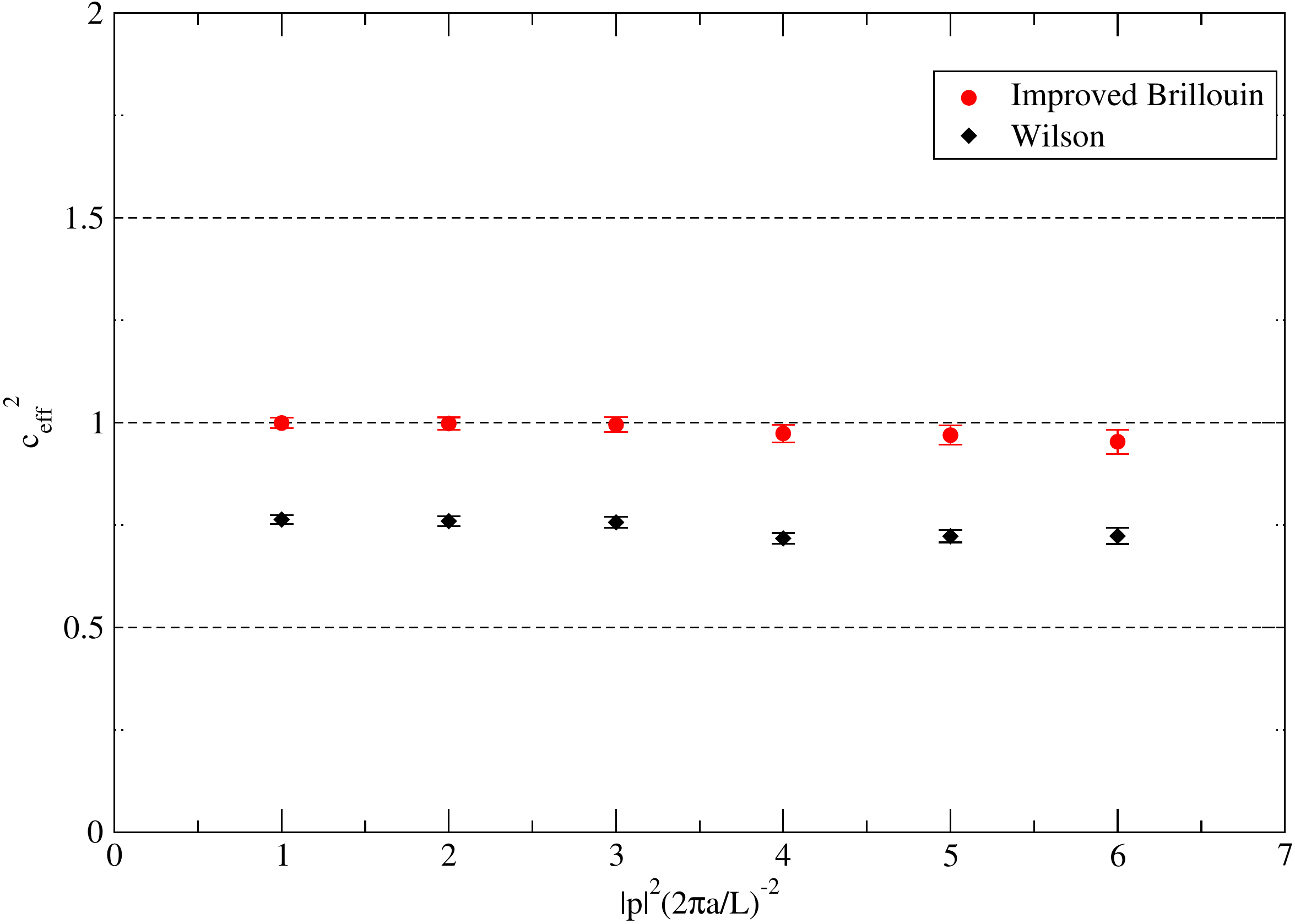}
    \includegraphics[scale=0.3,clip=on]{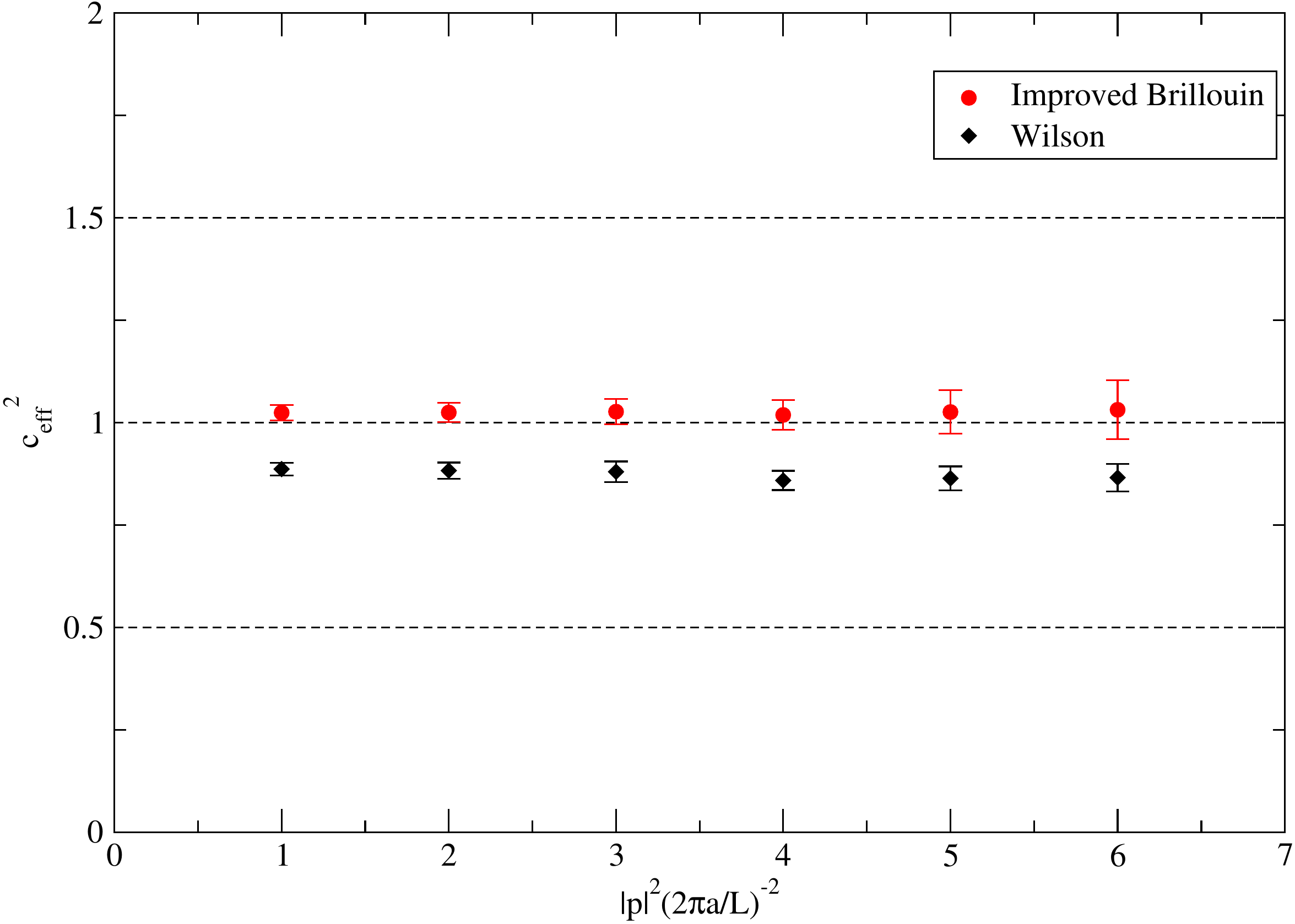}	
  \end{center}
  \caption{
    Effective speed of light as a function of normalized momentum
    squared at $a^{-1}=1.97$ GeV (left) and $a^{-1}=2.81$ GeV
    (right). 
    In each panel, data for improved Brillouin (filed circles) and
    Wilson (filed diamonds) fermions are plotted. 
  }
  \label{fig:spl}	
\end{figure}

\begin{figure}
  \begin{center}
    \includegraphics[width=9cm,clip=on]{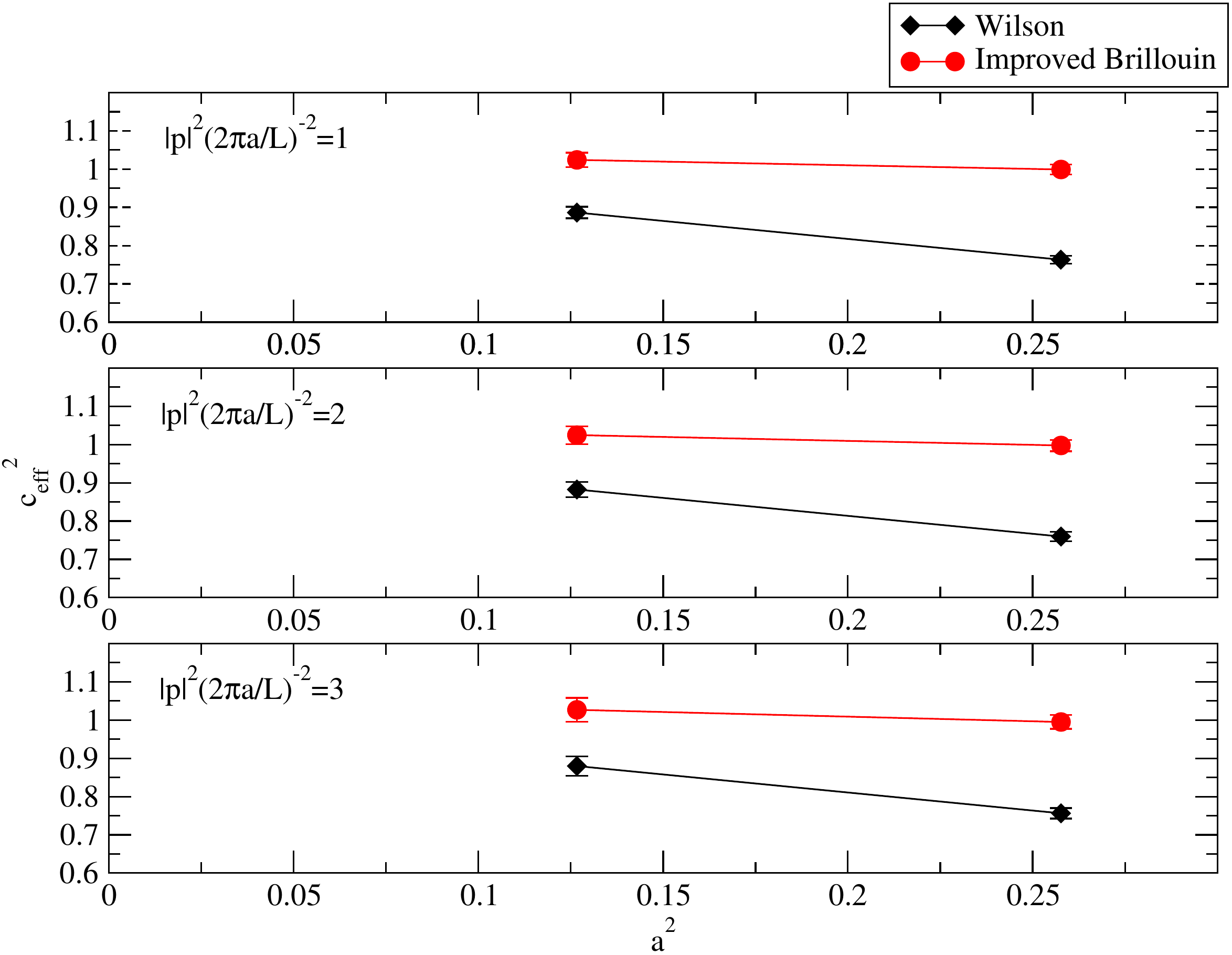}
  \end{center}
  \caption{
    Scaling of $c_{\rm eff}^{2}$ calculated at $|\vec{p}|^2$ =
    1 (upper), 2 (middle) and 3 (lower) in units of $(2\pi/L)^2$.
    In each panel, data for improved Brillouin (filed circles) and
    Wilson (filed diamonds) are plotted as a function of $a^2$
    [GeV$^{-2}$].
  } 
  \label{fig:sclspl}		
\end{figure}

The results are shown in Figure~\ref{fig:spl}.
Here $c_{\rm eff}^2$ is plotted against $|\vec{p}|^2$ for Wilson
(black) and the improved Brillouin (red) fermions.
Already at $1/a$ = 2.0~GeV (left) the dispersion relation of the 3-GeV
meson follows that of continuum theory, $c=1$, very precisely 
(within the statistical error) 
when the improved Brillouin fermion is employed.
With the Wilson fermion, the deviation is as large as 30\%.
Such large deviation is allowed in the effective theory approaches
\cite{ElKhadra:1996mp}, where the rest mass $m_1$ and the kinetic mass
$m_2$ are treated differently and only $m_2$ is taken as physical.
With the improved Brillouin fermion, this is not necessary.
Scaling towards the continuum limit is demonstrated in
Figure~\ref{fig:sclspl}.
In three panels, the results at normalized momentum squared are shown
as a function of $a^2$.
With the Brillouin fermion, we do not see any deviations from the
continuum at the level of 1\%, which is the size of the statistical error.

\begin{figure}
  \begin{center}
    \includegraphics[width=8cm,clip=on]{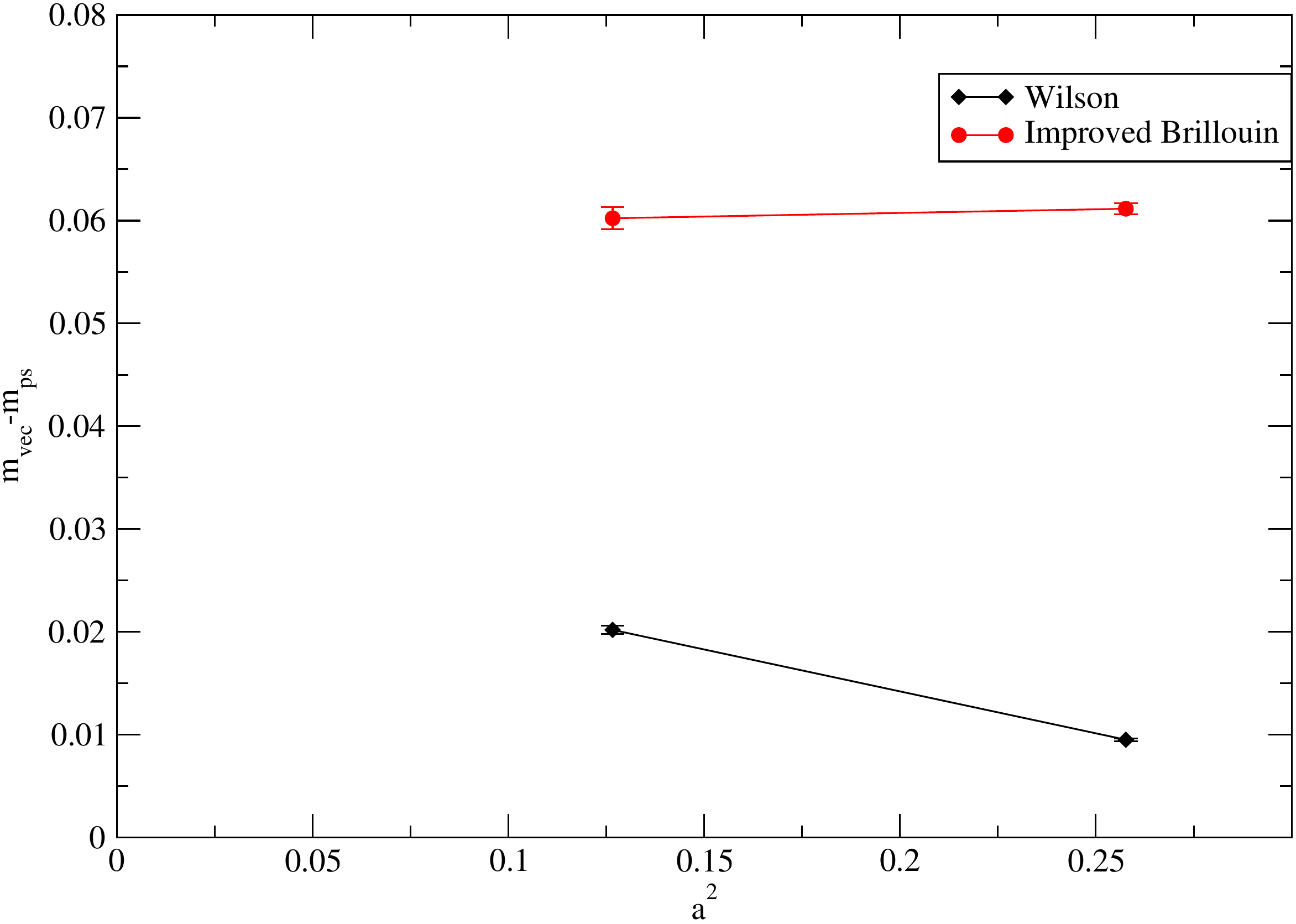}
  \end{center}
  \caption{
   Continuum scaling of the hyperfine splitting $m_{vec}-m_{ps}$
   [GeV].
   Results with the Wilson (black) and improved Brillouin (red)
   fermions are plotted as a function of $a^2$ [GeV$^{-2}$].
  }
  \label{fig:sclhyp}				
\end{figure}

In Figure~\ref{fig:sclhyp}, we show a similar scaling test of the two
formulations for the hyperfine splitting $m_{vec}-m_{ps}$ of the 3-GeV
heavy-heavy meson.
Also for this quantity, the scaling towards the continuum limit is
much better with the improved Brillouin fermion.

\begin{figure}
  \begin{center}
    \includegraphics[width=8cm,clip=on]{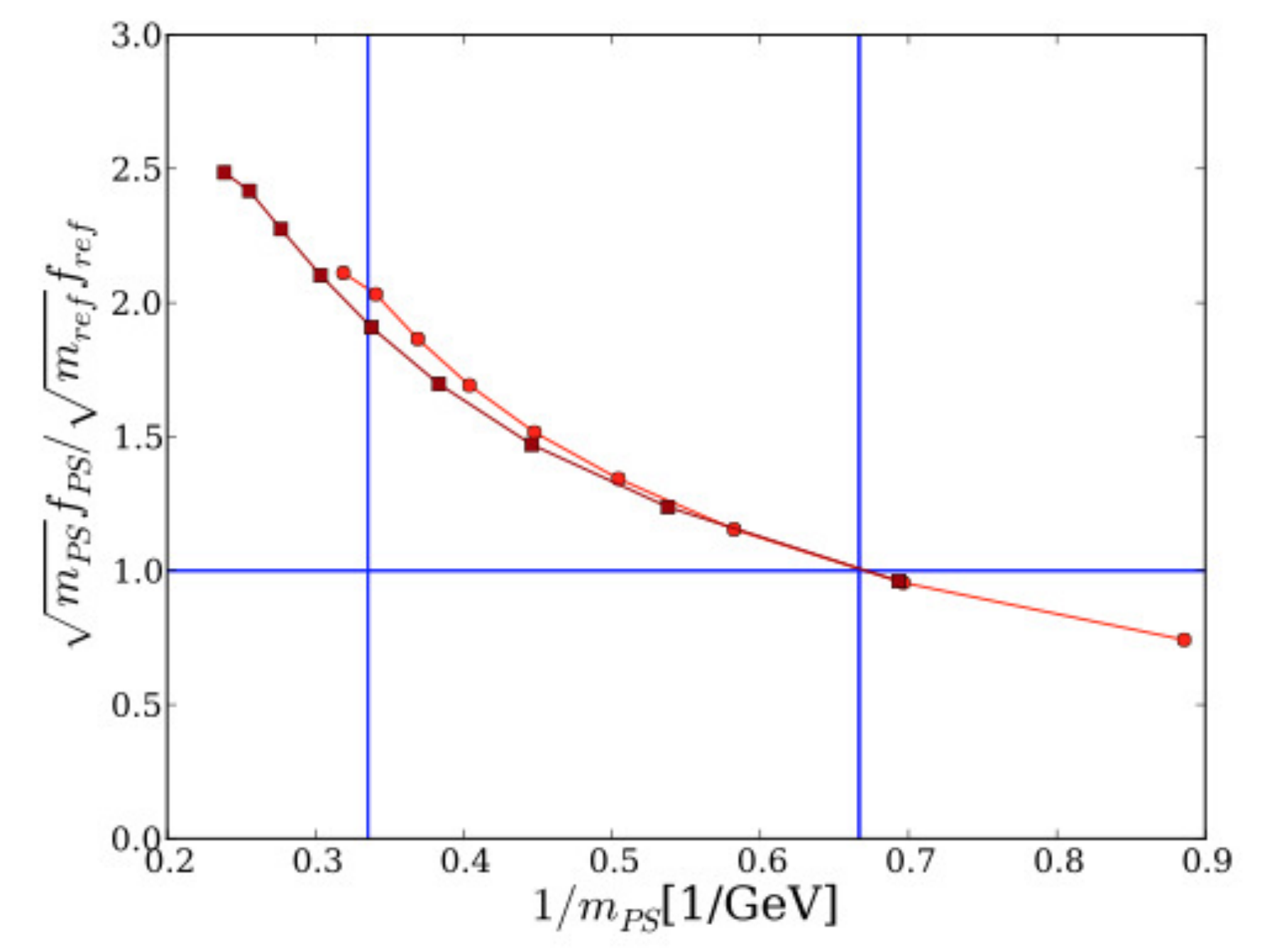}
  \end{center}
  \caption{
    Heavy-heavy pseudo-scalar meson decay constants calculated at two
    lattice spacings, 2.0~GeV (red) and 2.8~GeV (black),
    with the domain-wall fermion.
    The results for $f_{PS}\sqrt{m_{PS}}$ are normalized by the value at
    $m_{PS}$ = 1.5~GeV and plotted as a function of $1/m_{PS}$.
  }
  \label{fig:hhdc}
\end{figure}

Finally, we briefly describe a calculation of the heavy-heavy decay
constant using the domain-wall fermion.
It is well known that the domain-wall fermion mechanism breaks down at large
$am$ ($\simeq 0.5$)
\cite{Jansen:1992tw,Golterman:1992ub,Christ:2004gc},
but the real question is where it shows up in numerical calculations.
In Figure~\ref{fig:hhdc} we plot the decay constant
$f_{PS}\sqrt{m_{PS}}$ as a function of $1/m_{PS}$.
The data are normalized by the value at $m_{PS}$ = 1.5~GeV, so that
the renormalization constant cancels out.
We observe a good indication of scaling of $f_{ps}$ with the pseudo-scalar mass $m_{ps} \sim 3 GeV$, 
which suggests that the lattices of $1/a = 2-4 GeV$ could be used 
for the direct extraction of the properties of D mesons.
A complete continuum limit study of this and alternative heavy quark discretizations is needed 
to come to a final conclusion on this matter.

\section{Summary and plans}
Relativistic formulation for heavy quark has an advantage that no
tuning of parameters depending on the heavy quark mass is necessary.
Therefore, as fine-lattice dynamical QCD simulations has become
realistic, such a brute-force approach could be a powerful alternative
to the effective theory approaches, 
provided that the possible $(am)^n$ corrections are under control.

We are performing various scaling tests of relativistic formulations
on quenched lattices, and so far have obtained promising results.
In the future we plan to extend the study by adding finer lattices and
more choices of fermion formulations.

\vspace*{1cm}
Numerical simulations are performed on the IBM System Blue Gene
Solution (Blue Gene /Q) at High Energy Accelerator Research
Organization (KEK) under a support of its Large Scale Simulation
Program (No.~12/13-04). 
This work is supported in part by the Grant-in-Aid of the Japanese
Ministry of Education (No. 21674002) and the 
SPIRE (Strategic Program for Innovative Research) Field5 project. 
The research leading to these results has also received funding from the European Research Council under
the European Community's Seventh Framework Programme (FP7/2007-2013) ERC grant
agreement No 279757.

\newcommand{\J}[4]{{#1} {\bf #2} (#3) #4}
\newcommand{\RMP}{Rev.~Mod.~Phys.}
\newcommand{\MPL}{Mod.~Phys.~Lett.}
\newcommand{\IJMP}{Int.~J.~Mod.~Phys.}
\newcommand{\NP}{Nucl.~Phys.}
\newcommand{\NPSup}{Nucl.~Phys.~{\bf B} (Proc.~Suppl.)}
\newcommand{\PL}{Phys.~Lett.}
\newcommand{\PRD}{Phys.~Rev.~D}
\newcommand{\PRL}{Phys.~Rev.~Lett.}
\newcommand{\AP}{Ann.~Phys.}
\newcommand{\CMP}{Commun.~Math.~Phys.}
\newcommand{\CPC}{Comp.~Phys.~Comm.}
\newcommand{\PTP}{Prog. Theor. Phys.}
\newcommand{\Suppl}{Prog. Theor. Phys. Suppl.}
\newcommand{\JHEP}{JHEP}
\newcommand{\PoS}{PoS}

\end{document}